\documentclass[preprint,
superscriptaddress,
 amsmath,amssymb,
 aps, physrev,
]{revtex4-2}
\usepackage{amssymb}
\usepackage{cancel}
\usepackage[dvipsnames]{xcolor}
\definecolor{mygreen}{rgb}{0.03, 0.8, 0.39}
\usepackage{graphicx}
\usepackage[colorlinks=true, allcolors=blue]{hyperref}
\usepackage{dcolumn}
\usepackage{bm}

\begin{document}

\title{Nonlinear interface effects in multilayered structures: vibro-acoustic modeling and experimental analysis}

\author{Antoine DEMIQUEL}
\email{antoine.demiquel@insa-lyon.fr}
\affiliation{INSA Lyon, LVA, UR677, 69621 Villeurbanne, France}
\affiliation{ENTPE, Ecole Centrale de Lyon, CNRS, LTDS, UMR5513, 69518 Vaulx-en-Velin, France}
\affiliation{LabEx CeLyA - Centre Lyonnais d’Acoustique,
36 avenue Guy de Collongue, 69134 Ecully, France}

\author{Kerem EGE}
\email{kerem.ege@insa-lyon.fr}
\affiliation{INSA Lyon, LVA, UR677, 69621 Villeurbanne, France}
\affiliation{LabEx CeLyA - Centre Lyonnais d’Acoustique,
36 avenue Guy de Collongue, 69134 Ecully, France}

\author{Emmanuel GOURDON}
\email{emmanuel.gourdon@entpe.fr}
\affiliation{ENTPE, Ecole Centrale de Lyon, CNRS, LTDS, UMR5513, 69518 Vaulx-en-Velin, France}
\affiliation{LabEx CeLyA - Centre Lyonnais d’Acoustique,
36 avenue Guy de Collongue, 69134 Ecully, France}

\date{\today}
\begin{abstract}
This paper presents an experimental and theoretical study of the nonlinear behavior of imperfect interfaces in multilayer structures using an equivalent vibro-acoustic approach. The multilayer system is modeled through a Zig-Zag formulation, in which interfacial coupling conditions, stress continuity and displacement discontinuity, relate the kinematics of adjacent layers while preserving an independent description of each layer. This framework significantly reduces the number of kinematic unknowns without compromising the model accuracy.

An equivalent Kirchhoff–Love plate formulation is then introduced to derive a frequency-dependent bending stiffness representative of the global structural response. Experimental measurements of the transverse displacement field are performed using laser vibrometry and processed via the Corrected Force Analysis Technique (CFAT). The results demonstrate that the dynamic response of a three-layer beam with imperfect interfaces depends on the excitation level. In particular, variations in the equivalent bending stiffness are observed, revealing the nonlinear nature of the interfacial behavior. The proposed methodology is applied to a glass/epoxy/glass multilayer beam under various excitation levels.
\end{abstract}

\maketitle

\section{INTRODUCTION}
Multilayered structures, including laminated and sandwich structures are widely used for a broad range of applications~\cite{birman_review_2018}, such as transport~\cite{rao_recent_2003}, civil engineering~\cite{caniato_acoustic_2017}, microelectronics~\cite{zheng_research_2024}, and biomedical engineering~\cite{manteghi_investigation_2017}. In the vibroacoustic domain, these structures offer attractive combinations of stiffness, mass, and damping, while allowing different materials and functions to be integrated within a single component. Recent advances in architected and adaptive materials have further extended the potential of multilayered systems. In such structures, the geometry and active transducers are intentionally designed to achieve tunable and programmable mechanical properties, such as stiffness modulation, wave control, or vibration attenuation \cite{collet_adaptive_2014,da_s_raqueti_equivalent_2025,zeng_smart_2022}. 

The dynamic behavior of such structures has been extensively investigated, leading to a wide range of modeling approaches \cite{carrera_assessment_2000}. Among them, the formulation proposed by Guyader~\textit{et~al.}~\cite{guyader_acoustic_1978, Guyader2007Viscoelastic} describes each layer using independent kinematic fields, while interfacial coupling conditions ensure continuity and reduce the number of degrees of freedom. Within this framework, the global behavior of the sandwich structure is thus expressed through a reference layer from which the kinematics of all other layers are derived. This approach also enables the definition of equivalent mechanical properties of the multilayered by an equivalent thin plate representation. Building on this foundation, Marchetti~\textit{et~al.}~\cite{marchetti_structural_2020} extended Guyader’s work to panel structures with anisotropic layers, while providing a detailed implementation of the model. 

Despite these advances, the dynamic response of multilayered systems remains strongly influenced by the behavior of the interfaces between layers \cite{adekola_partial_1968,di_sciuva_geometrically_1997,cheng_theory_1996}. Moreover, recent works in multi-component metamaterials show that interfacial constraints are crucial for vibration damping \cite{perez_ramirez_effective_2024,hermann_design_2024,hermann_unveiling_2026}. In practice, bonding is rarely perfect: defects in adhesive, partial debonding, thin compliant interlayers or progressive damage give rise to imperfect interfaces. These imperfections can significantly affect the structural response, including bending stiffness, natural frequencies, damping, and vibro-acoustic performance \cite{auquier_imperfect_2024,massabo_efficient_2014,wang_review_2025}. A common way to model these interfaces is to represent them as sliding or spring-type interfaces, where the interfacial tractions are related to the relative  displacements between adjacent layers. Within this framework, several equivalent models have been proposed to capture the influence of interfacial compliance. These models describe its effect on the effective bending stiffness and vibration characteristics of multilayers, often revealing a strong nonlinear dependence of the apparent stiffness on the interface parameters \cite{wang_review_2025,massabo_efficient_2014,massabo_assessment_2015}. 

More recently, the Guyader formulation has been extended to account for imperfect interfaces through the introduction of an equivalent interface parameter \cite{auquier_equivalent_2022,auquier_imperfect_2024}. These studies showed that the quality of the interlayer coupling induces a shift of the low-frequency dynamic characteristics, with more imperfect interfaces leading to larger frequency shifts. However, the extension of this approach to  nonlinear interface laws remains an open question, which constitutes the main motivation of the present work.

The objective of this study is to develop an experimental methodology for characterizing the nonlinear behavior of multilayer structures through the identification of an equivalent amplitude-dependent interface parameter. This approach relies on displacement field measurements performed under varying excitation levels. To this end, we employ the model developed by Arasan~\textit{et~al.}~\cite{arasan_simple_2021-1}, which describes the dynamic evolution of the bending stiffness using an explicit sigmoid function. By analyzing the variation of the sigmoid parameters induced by nonlinear effects, it becomes possible to identify and characterize an amplitude-dependent effective interface parameter representative of the nonlinear interfacial behavior.

\section{DYNAMIC MODELING OF MULTILAYER STRUCTURES WITH IMPERFECT INTERFACES}
In this section, we outline the Zig-Zag multilayer model developed by Guyader and Lesueur~\cite{guyader_acoustic_1978}, incorporating imperfect interface conditions introduced by Auquier~\textit{et al.}~\cite{auquier_equivalent_2022}, which are employed in this work. This class of models \cite{lekhnitskii1935strength,Ren1986ANT,sun_theories_1973} assigns independent kinematic fields to each layer while applying interface conditions laws that couple the layers and limit the number of degrees of freedom. The global behavior of the sandwich structure is thus expressed through a reference layer, from which the kinematics of all other layers are derived. The equations of motion are subsequently obtained using Hamilton’s principle, combining the kinetic and strain energies of the system. A plane-wave assumption is then introduced to solve these equations and extract the dispersion curves of the structure. Finally, the multilayer system is compared, at each frequency, with an equivalent homogeneous Love–Kirchhoff plate \cite{love1888vibrations} whose bending rigidities define the dynamic effective properties of the multilayer.
\subsection{Presentation of the system and governing equations}
\label{Presentation_system}
\begin{figure*}[ht!]
\centering
\includegraphics[width=0.9\linewidth]{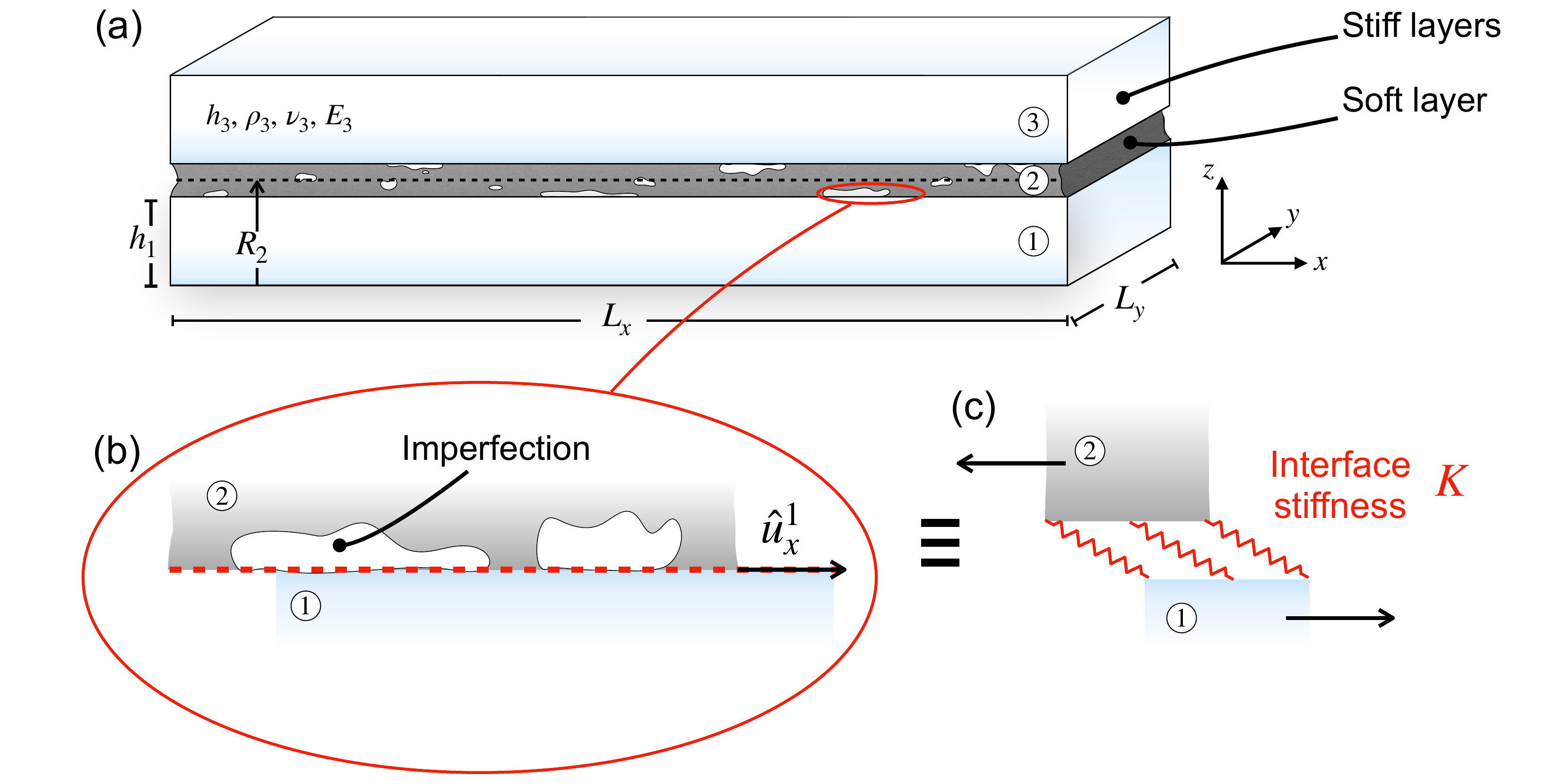}
\vspace{-0.5cm}
\caption{(a) Outline of the considered three-layered structure composed of stiff skin layers (light blue areas) and a soft core (gray area). $L_x$ ($L_y$) denotes the length (width) along the $x$-axis ($y$-axis) of the structure. (b) Imperfections in the core layer at the interfaces induce a sliding effect quantified by a jump function $\hat{u}_{\{x,y\}}$. The imperfect interfaces between the core and the bottom/top skin (red dashed line) are modeled using constitutive equations based on a spring-layer approach, connecting displacements and shear stresses at the interface linearly to an equivalent interface stiffness $K$ (cf.~Eq.~(\ref{traction_law_K})) or to an equivalent interface compliance $B$ (cf.~Eq.~(\ref{traction_law_B})).}
 \label{System}
\end{figure*}

The system studied experimentally and illustrated in Fig.~\ref{System}(a), is a three-layer structure ($N=3$) composed of rigid glass skins and a softer core made of two-component epoxy adhesive DP-190 3M, which serves both as a bonding agent and as an intermediate structural layer. Each of these layers is characterized by a thickness $h_n$, density $\rho_n$, Poisson’s ratio $\nu_n$, Young's modulus $E_n$. The corresponding material and geometric properties of the experimental sample, summarized in Table~\ref{tab:material_properties}, are taken from previous experimental studies carried out by Auquier~\textit{et~al.}~\cite{auquier_imperfect_2024} on similar samples and are used in the subsequent numerical analysis.
\begin{table}[h!]
\centering
\begin{tabular}{|l|cccc|}
\hline
Layer [$n$] & $h_n$ [mm] & $\rho_n$ [kg.m$^{-3}$] & $\nu_n$ [-] & $E_n$ [GPa]  \\ 
\hline
Glass skins [1, 3] & 3 & 2700 & 0.33 & 71  \\ 
Epoxy core [2]   & 0.35 & 1300 & 0.3 & 1 \\ 
\hline
\end{tabular}
\vspace{0.3cm}
\caption{Material and geometrical properties of the three-layered experimental sample.}
\label{tab:material_properties}
\end{table}

The formulation introduced by Guyader and Lesueur \cite{guyader_acoustic_1978}, derived from Reissner–Mindlin plate theory \cite{Reissner1945,Mindlin1951}, incorporates bending, membrane and linear shear effects are accounted for each layer $n$. The displacement fields of the layer $n=\{1,2,3\}$ write,
\begin{equation}
 \left \{
    \begin{split}
        u^n_x &=\psi^n_x(x,y,t)-(z-R_n)\left[\frac{\partial W(x,y,t)}{\partial x}+\phi^n_x(x,y,t)\right]\\
        u^n_y &=\psi^n_y(x,y,t)-(z-R_n)\left[\frac{\partial W(x,y,t)}{\partial y}+\phi^n_y(x,y,t)\right]\\
        u^n_z &=W(x,y,t)\\
    \end{split}
    \right.~,
    \label{Champ_des_deplacements}
\end{equation}
where $W$ denotes the transverse displacement, $\psi_x^n$ and $\psi_y^n$ correspond to the membrane displacement along the $x$ and $y$ directions, respectively. Finally, $\phi_x^n$ and $\phi_y^n$ are the rotations around the $Ox$ and $Oy$ axes. The length $R_n$ represents the location of the mid-surface of layer $n$ along the thickness coordinate $z$, defined with respect to the global reference axis.

Before proceeding with the development, it is crucial to keep in mind several key points about Eq.~(\ref{Champ_des_deplacements}). When employing this displacement field, it is essential to recognize the underlying assumptions, as they introduce inherent limitations to the model and define the range of its applicability. First, the model assumes a uniform transverse displacement $W$ accross the thickness of each layer, which implies that deformations along the $z$-direction are neglected. Consequently, thickness vibration modes, such as breathing modes, are not described by this model. The frequency $f_{S0}$ of the first breathing mode can be estimated using an equivalent mass–spring–mass model in which the two outer layers are represented by masses $m_1, m_3$ while the core layer is modeled by a normal stiffness $k_2$ \cite{ege_assessment_2018}. Using the material and geometric properties listed in Tab.~\ref{tab:material_properties}, the resulting expression and numerical value of the first breathing mode frequency are the following, 
\begin{equation}
    f_{S0}= \frac{1}{2\pi}\sqrt{\frac{k_2(m_1+m_3)}{m_1m_3}}=\frac{1}{2\pi}\sqrt{\frac{E_2}{h_2}\frac{\rho_1h_1+\rho_3h_3}{\rho_1h_1\rho_3h_3}} \approx 134~\text{kHz}~.
    \label{Breathing_frequency}
\end{equation}
Finally, the linear variation of shear effects through the thickness, represented by $\left(z\phi^n_{\{x,y\}}\right)$ introduces limitations in the prediction of bending modes at higher frequencies. At these frequencies, bending motion becomes increasingly coupled with shear deformations, which are negligible at low frequencies but play a significant role in accurately capturing the structural response at higher frequencies. With these key points clarified, the next step consists in the development formulation that leads to the dispersion curves of the equivalent homogeneous system.

Eq.~(\ref{Champ_des_deplacements}) shows that all layers $n$ share the same expression of the displacement field. The global displacement fields of the multilayer system is then obtained by transmitting this formulation through the layers and across the interfaces using the appropriate interface relations. In practice, the intermediate soft bonding layer is not perfectly uniform, air bubbles and local material gaps can be present in the core layer and at the interfaces, as illustrated in Fig.~\ref{System}(a, b). These manufacturing-induced heterogeneities generate inhomogeneous interfaces with mechanical properties that are not fully controllable, allowing relative motion between layers, and resulting in discontinuities of the displacement field through the thickness.

To account for these effects, the interfacial behavior is modeled using a linear spring-layer approach, which couples the transverse displacements and shear stresses of adjacent layers, as shown in Fig.~\ref{System}(c).

\subsection{Imperfect interface relations: 
Piecewise linear approximation}
\label{Interface relations: 
Piecewise linear approximation}
The original multilayer formulation developed by Guyader and Lesueur \cite{guyader_acoustic_1978} assumed ideal, perfectly bonded interfaces. This framework is extended by Auquier~\textit{et~al.}~\cite{auquier_equivalent_2022} to incorporate the effects of imperfect interfaces. In this extended model, the influence of interface imperfections is introduced through jump functions $\hat{u}^{n-1}_{\{x,y\}}$ ($\hat{u}^{n-1}_x$ and $\hat{u}^{n-1}_y$  ), see Fig.~\ref{System}(b). These functions quantify the relative transverse displacements between layer $(n-1)$ layer and the overlying layer $n$  at their common interface located at $z_n$ along the $x$ and $y$ directions,
\begin{equation}
   \left. u_{\{x,y\}}^n\right\rvert_{z_n=R_n-\frac{h_n}{2}}=\left.u_{\{x,y\}}^{n-1}\right\rvert_{z_n=R_{n-1}+\frac{h_{n-1}}{2}}+ \hat{u}^{n-1}_{\{x,y\}}~,
    \label{displacements_conditions}
\end{equation}
with $\hat{u}^N_{\{x,y\}}=0$ since the index layer $n$ corresponds to the lower layer. For $n=N=3$ there is no interface.

Since physical contact between the layers is maintained at $z=z_n$, the continuity of the shear stresses across the interface is preserved,
\begin{equation}
    \left. \sigma_{\{x,y\}z}^n\right\rvert_{z_n=R_n-\frac{h_n}{2}}=\left.\sigma_{\{x,y\}z}^{n-1}\right\rvert_{z_n=R_{n-1}+\frac{h_{n-1}}{2}}=\hat{\sigma}^{n-1}_{\{x,y\}z}~.
    \label{shear_stresses_conditions}
\end{equation}

Eqs.~(\ref{displacements_conditions}–\ref{shear_stresses_conditions}) describe the governing equations relations that characterize the interaction between neighboring layers. A classical approach used to link the relative transverse displacements to the shear-stresses at an interface $z_n$ is to use linear shear–slip traction laws \cite{adekola_partial_1968,di_sciuva_geometrically_1997,cheng_theory_1996}, see Fig.~\ref{System}(c). This approach provides a simplified representation of the interface deformation resulting from relative motion of the layers. The constitutive relations of
the linear and uncoupled interface given in Eq.~(\ref{traction_law}) correspond to the simplest admissible model for delamination \cite{williams_general_1997},
\begin{subequations}
    \begin{equation}
        \left.\hat{\sigma}^n_{\{x,y\}z}\right\rvert_{z=z_n}= K_{\{x,y\}}^n\left.\hat{u}^n_{\{x,y\}}\right\rvert_{z=z_n}~, \hspace{1cm}\text{with}\hspace{0.5cm} n=\{1,2,\ldots,N-1\}~,
        \label{traction_law_K}
    \end{equation}
    or
    \begin{equation}
         \left.\hat{u}^n_{\{x,y\}}\right\rvert_{z=z_n} = B_{\{x,y\}}^n\left.\hat{\sigma}^n_{\{x,y\}z}\right\rvert_{z=z_n}~, \hspace{1cm}\text{with}\hspace{0.5cm} n=\
         \{1,2,\ldots,N-1\}~.
         \label{traction_law_B}
    \end{equation}
    \label{traction_law}
\end{subequations}
The parameters $K^n$ and $B^n$ represent the tangential stiffness and the corresponding compliance of the interface. In reality, the mechanical response of interfaces is often much more complex and typically exhibits nonlinear behavior. Such nonlinearities may originate from several physical mechanisms, including the elastic response of thin bonding layers, cohesive or bridging effects due to translaminar reinforcements, progressive material degradation and failure, or elastic contact along partially delaminated surfaces, as discussed by Massab\`o and Campi~\cite{massabo_efficient_2014}.
\begin{figure}[ht!]
    \centering
    \includegraphics[width=0.95\linewidth]{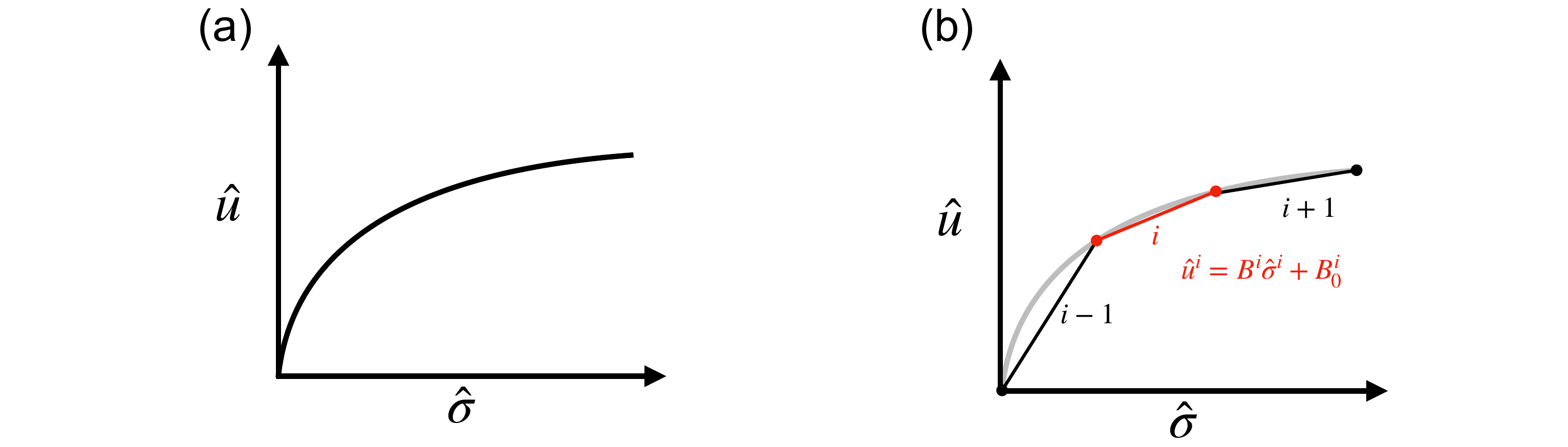}
    \vspace{-0.5cm}
    \caption{(a) Example of interface behavior characterized by a nonlinear interface constitutive relation and (b) its approximation using a piecewise linear representation.}
    \label{fig:piecewise}
\end{figure}

For computational purposes, nonlinear traction laws, such as the one depicted in Fig.~\ref{fig:piecewise}(a), are commonly approximated by piecewise linear relations as shown in panel~(b). Each segment $i$ is described by an affine function linking relative displacements to shear stresses, 
\begin{equation}
    \hat{u} = \sum_i \left(B^i\hat{\sigma}^i+B_0^i\right)~.
    \label{piecewise}
\end{equation}
In this study, interfaces are characterized using a single scalar equivalent parameter $B$, assumed homogeneous and isotropic. Specifically, for a given interface $n$, we assume ${B_{\{x,y\}}^n = B}$ and ${B_{0_{\{x,y\}}}^n=B_0}$ at all points of the interface along the $x$ and $y$ directions. When ${B=0}$~m.Pa$^{-1}$, the interface is considered as perfect, the two adjacent layers are perfectly bounded. Conversely, as $B\rightarrow \infty$, the interface is like "non-existent", representing complete debonding between the layers.

Collectively, the interface relations Eqs.~(\ref{displacements_conditions}, \ref{shear_stresses_conditions}, \ref{traction_law}, \ref{piecewise}) allow the kinematic variables of the layer  $n$, collected in the vector $\{L^{n}\}$, to be expressed in terms of those of the preceding layer $n-1$,$\{L^{n-1}\}$, via the transfer matrix $[T^n]$,
\begin{equation}
\underbrace{
\begin{pmatrix}
      \frac{\partial W}{\partial x} \\
      \phi^n_x \\
      \psi^n_x \\
      \frac{\partial W}{\partial y} \\
      \phi^n_y \\
      \psi^n_y
\end{pmatrix}
}_{\{L^n\}}
=
\underbrace{
\begin{pmatrix}
    1 & 0 & 0 & 0 & 0 & 0  \\
    0 & \mathcal{A}^n_{x} & 0 & 0 & 0 & 0  \\
    \mathcal{B}^n & \mathcal{C}^n_{x} & 1 & 0 & 0  & 0  \\
    0 & 0 & 0 & 1 & 0 & 0  \\
    0 & 0 & 0 & 0 & \mathcal{A}^n_{y}  & 0   \\
    0 & 0 & 0 & \mathcal{B}^n & \mathcal{C}^n_{y} & 1
\end{pmatrix}
}_{[T^n]}
\underbrace{
\begin{pmatrix}
      \frac{\partial W}{\partial x} \\
      \phi^{n-1}_x \\
      \psi^{n-1}_x \\
      \frac{\partial W}{\partial y} \\
      \phi^{n-1}_y \\
      \psi^{n-1}_y
\end{pmatrix}
}_{\{L^{n-1}\}}-\underbrace{
\begin{pmatrix}
      0\\
      0 \\
      B_0 \\
      0 \\
      0 \\
      B_0
\end{pmatrix}
}_{\{B_0\}}~, \hspace{1cm}\text{with}\hspace{0.5cm} n=\{2,\ldots,N\}~.
\label{Matrix_T}
\end{equation}

The coefficients $\mathcal{A}^n$, $\mathcal{B}^n$, and $\mathcal{C}^n$ composing the transfer matrix $[T^n]$ are provided in Appendix \ref{Appendix_A}. By recursively applying Eq.~(\ref{Matrix_T})
 from the outermost layer $N$ down to the first layer, 
 \begin{equation}
     \{L^n\}=[T^n][T^{n-1}][T^{n-2}]\cdots[T^2]\{L^{1}\}-(n-1)\{B_0\}~, \hspace{1cm}\text{with}\hspace{0.5cm} n=\{2,\ldots,N\}~,
 \end{equation}
 the displacement of each layer can be expressed in terms  of the kinematic variables of the first layer, yielding
  \begin{equation}
      \left \{
      \begin{split}
      u^n_x& = \psi^1_x(x,y,t)+F_\omega \frac{\partial W(x,y,t)}{\partial x}+F^n_{x}\phi^1_x(x,y,t)-(n-1)B_0\\ 
      u^n_y& =\psi^1_y(x,y,t)+F_\omega \frac{\partial W(x,y,t)}{\partial y}+F^n_{y}\phi^1_y(x,y,t)-(n-1)B_0\\ 
      u^n_z& =W(x,y,t) \\ 
            \end{split}
      \right.~,
      \label{champ_depl_n_fonction_1}
  \end{equation}
with 
\begin{equation}
    F^n_{\{x,y\}}=\alpha^n_{\{x,y\}}(R_n-z)+\gamma^n_{\{x,y\}}~, \hspace{1cm} \hspace{1cm} \text{and}\hspace{1cm} F_\omega = R_1-z~.
    \label{coeff_F}
\end{equation}

For isotropic layers, the $x$- and 
$y$- components of the displacement field are uncoupled. The coefficients $F^n_{\{x,y\}}$, defined in Eq.~(\ref{coeff_F}), depend on the quantities  $\alpha^n_{x}, \alpha^n_{y}, \gamma^n_x, \gamma^n_y$ and $\beta^n$, which are expressed in Appendix~\ref{Appendix_A} cf.~Eqs.~(\ref{betan}-\ref{gammany}), as functions of the geometric and mechanical properties of each layer. It is noteworthy that the interface parameter $B$ influences only the  $\gamma^n_{\{x,y\}}$ coefficients, while a nonzero value of $B_0$ represents a displacement discontinuity at the interfaces between layers.

\subsection{Equivalent bending
stiffness}
Once the displacement field of the $n$-th layer has been expressed relative to the reference layer, the equivalent dynamic properties of the multilayer structure can be determined. To this end, the sandwich medium is replaced by an equivalent thin plate model, allowing the multilayer, at a given angular frequency $\omega=2\pi f$, to be represented by a single plate characterized with effective dynamic parameters. This approximation relies on the assumption that the transverse displacement remains uniform across the plate thickness. The model used is the classical Love–Kirchhoff thin plate theory~\cite{love1888vibrations}, which yields the following expression for the effective bending stiffness,
\begin{equation}
    D = \frac{\rho h \omega^2}{k_b^4}~.
    \label{Kirchhoff}
\end{equation}
In this relation, $h=\sum_{n=1}^3 h_n$ is the total thickness of the structure, obtain as the sum of individual layer thickness $h_n$ . The equivalent mass density is defined as $\rho = \sum_{n=1}^3\frac{\rho_n h_n}{h}$, where $\rho_n$ represents the density of the $n_{th}$ layer.$\omega$ is the angular frequency, while $k_b$ corresponds to the bending wavenumber, which is extracted from the dispersion curves. These dispersion relations are derived by substituting a plane-wave solution,
\begin{equation}
  \begin{pmatrix}
      W \\
      \psi_x \\
      \phi_x \\
      \psi_y \\
      \phi_y 
\end{pmatrix}  = \begin{pmatrix}
      A \\
      B \\
      C \\
      D \\
      E
\end{pmatrix}e^{-jkx}e^{j\omega t}=\bold{W_0}e^{-jkx}e^{j\omega t}~,
\end{equation}
into the equations of motion, obtained from the Euler-Lagrange formalism expressed in differential form for each generalized coordinate. As the derivation follows established procedures available in the literature \cite{Guyader2007Viscoelastic,marchetti_structural_2020,auquier_equivalent_2022}, only the final result is reported here, leading to the following eigenvalue problem,
\begin{equation}
    (\bold{K}-\omega^2\bold{M})\bold{W_0} = \bold{0}~.
\label{EVP}
\end{equation}

The dispersion relations curves are obtained by solving the characteristic equation of the eigenvalue problem written in Eq.~(\ref{EVP}) where the matrices $\textbf{K}$ and $\textbf{M}$ represent the contribution of the strain energy and inertia effects. These matrices depend on a set of coefficients $\lambda_i$ and $\delta_i$, whose explicit expressions are not recalled here but can be found in Marchetti~\textit{et~al.}~\cite{marchetti_structural_2020}. Assuming that the total thickness of the multilayer plate is small, inertia contributions related to membrane deformation, transverse shear, and rotational motion can be neglected. Under this assumption, all inertia coefficients vanish except for $\delta_{13}$. Consequently, Eq.~(\ref{EVP}) reduces significantly and, after substitution, leads to a single governing equation describing the transverse motion with amplitude $A$. By combining this relation with Eq.~(\ref{Kirchhoff}), an expression for the effective bending stiffness is obtain. This formulation is independent of the bending wavenumber $k_b$ and describes its frequency-dependent behavior,
\begin{equation}
    A_4D^{3/2}+\tilde{A}_3D-A_1A_4D^{1/2}-A_1\tilde{A}_3+\tilde{A}_2=0~,
    \label{Bending_stifness_Guyader}
\end{equation}
with,
$A_1=\lambda_1-\frac{\lambda_5^2}{\lambda_3}$, $\tilde{A}_2=\omega\sqrt{\rho h}\left(\tilde{\lambda}_4-\frac{\lambda_5\tilde{\lambda}_6}{\lambda_3}\right)^2$, $\tilde{A}_3=\omega\sqrt{\rho h}\left(\tilde{\lambda}_2-\frac{\tilde{\lambda}_6^2}{\lambda_3}\right)^2$, $A_4=\lambda_{37}$. The coefficients with a tilde "$\sim$" denote terms that explicitly depend on the equivalent interface parameter $B$, which appears in the expressions of $\tilde{\lambda_i}$ for $i =\{2, 4, 6\}$. 

The multilayer model having been introduced along with the influence of imperfect interface conditions, the next section focus on the frequency-dependent evolution of the effective bending stiffness.

\section{ASYMPTOTIC ANALYSIS AND SIGMOID-BASED MODEL TO OBTAIN AN EXPLICIT EXPRESSION OF $D(f)$}
\label{sec:sigmoid}
 \subsection{Asymptotic analysis}
 According to Eq.~(\ref{Bending_stifness_Guyader}), the effective bending stifness exhibits a sigmoidal evolution, corresponding to a smooth transition between two asymptotic regimes associated with low and high frequencies  , see Fig.~\ref{Guyader_Arasan}(a). The expressions of these asymptotic limits are derived below. The frequency dependence arises through the coefficients $\tilde{A}_2$ and $\tilde{A}_3$, whereas $A_1$ and $A_4$ remain frequency independent. 
 
 In the low frequency limit ($\omega\rightarrow 0$), the frequency-dependent contributions vanish, and Eq.~(\ref{Bending_stifness_Guyader}) reduces to,
 \begin{equation*}
     A_4D_{low}^{3/2}-A_1A_4D_{low}^{1/2}=0~,
 \end{equation*}
 which leads to the asymptotic value,
\begin{equation}
    D_{low} = A_1~.
    \label{D_low}
\end{equation}
In the high frequency limit ($\omega\rightarrow \infty$), $\tilde{A}_2$ and $\tilde{A}_3$ dominate, and Eq.~(\ref{Bending_stifness_Guyader}) simplifies to,
 \begin{equation*}
     \tilde{A}_3D_{high}-A_1\tilde{A}_3+\tilde{A}_2 \approx 0~,
 \end{equation*}
 yielding the high-frequency asymptotic value, which depends on the interface parameter $B$,
\begin{equation}
    \tilde{D}_{high} = A_1-\frac{\tilde{A}_2}{\tilde{A}_3}~.
    \label{D_high}
\end{equation}

At this point, it is important to note that, within the proposed model, the low-frequency bending stiffness $D_{low}$ is not affected by the quality of the interface coupling, whereas the high-frequency bending stiffness $\tilde{D}_{high}$ is. Since the parameter $B$ enters the Eq.~(\ref{Bending_stifness_Guyader}) through the coefficients $\tilde{\lambda}_2$, $\tilde{\lambda}_4$ and $\tilde{\lambda}_6$, the bending wavenumber $k_b$ is directly affected by the interfaces behavior. Consequently, any nonlinear mechanism occurring at the interfaces results in a shift in the dispersion branches. Beyond the identification of such nonlinear effects, the present work aims to quantify the evolution of the interface parameter as a function of the excitation amplitude applied to the sample. For this purpose, a sigmoid-based model is employed.

 \subsection{Sigmoid-based model}
\begin{figure}[ht!]
\centering
\includegraphics[width=0.9\linewidth]{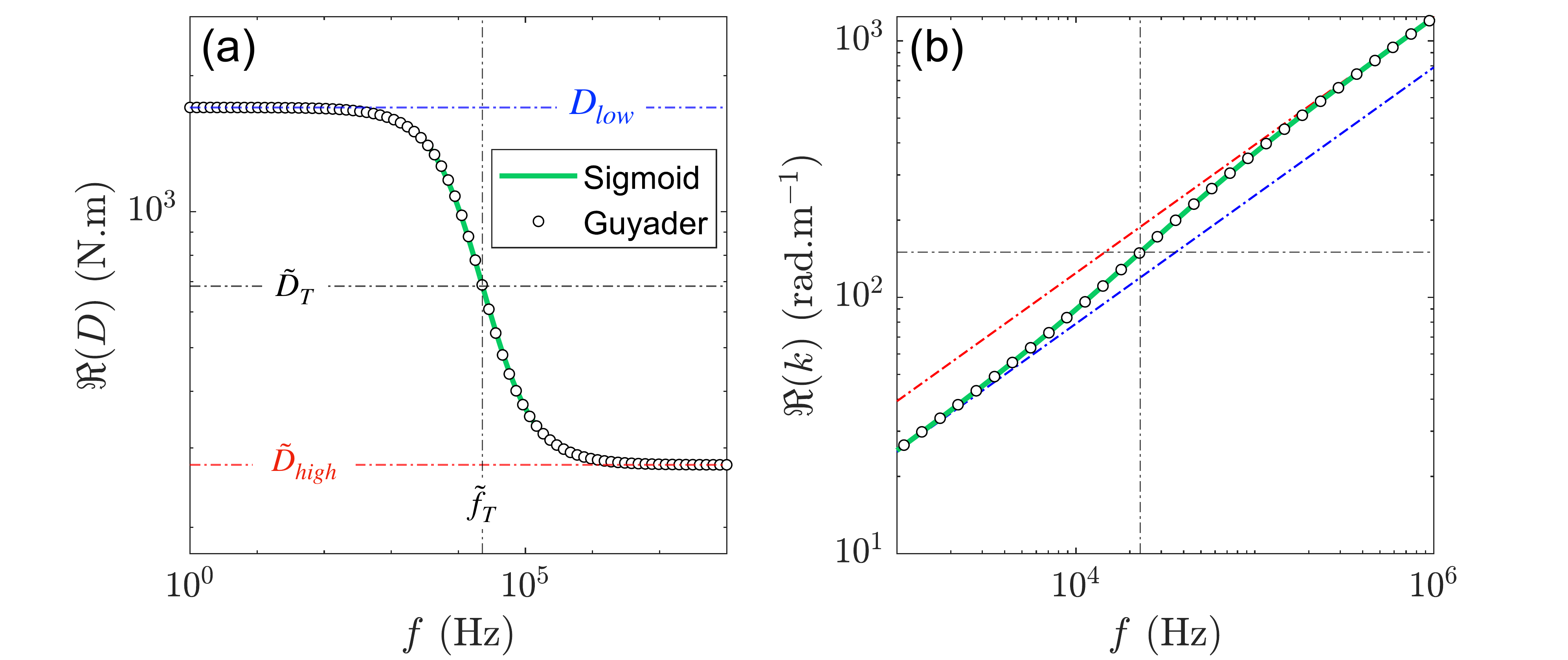}
\vspace{-0.5cm}
\caption{(a) Equivalent bending stiffness and (b) equivalent wavenumbers obtained from the proposed sigmoid model (\textcolor{mygreen}{\rule[0.5ex]{1.5em}{1pt}}) cf. Eq.~(\ref{Sigmoid_Arasan}). Results are compared with those obtained from the Guyader model ($\circ$) cf. Eq.~(\ref{Kirchhoff}).} 
 \label{Guyader_Arasan}
\end{figure}

This section presents a sigmoid-based model that provides an explicit description of the frequency dependence of the bending stiffness $D(f)$ predicted by the Guyader model Eqs.~(\ref{Kirchhoff}, \ref{Bending_stifness_Guyader}). This explicit formulation is particulary useful for isolating the bending wave branch from experimentally measured dispersion curves  when the interfaces properties are not known, see Sec.~\ref{bending_k_isolation}. To this end, the sigmoid model introduced by Arasan~\textit{et~al.}~\cite{arasan_simple_2021-1} is adopted, leading to the following expression for the equivalent bending stiffness,
\begin{equation}
    \text{log}_{10} \tilde{D}_\text{sig}(f)= \frac{f_T^R \text{log}_{10} D_\text{low}+f^R \text{log}_{10} \tilde{D}_\text{high}}{f^R+f_T^R} ~,
    \label{Sigmoid_Arasan}
\end{equation}
where $D_{low}$, $\tilde{D}_{high}$ are the dynamic bending stiffness asymptote of low and high-frequencies Eqs.~(\ref{D_low}, \ref{D_high}), while $f_T$ and $R$ control the transition frequency and the slope of the curve at the inflection point ($f_T$, $D_T$). From Eq.~(\ref{Sigmoid_Arasan}), the curvature of $\tilde{D}_{sig}$ changes sign at the midpoint, so
\begin{equation}
   \tilde{D}_T = \sqrt{D_{low}\tilde{D}_{high}}~.
\end{equation}
The transition frequency can be obtained by evaluating Eq.~(\ref{Bending_stifness_Guyader}) at $D=\tilde{D}_T$ and $f=f_T$ leading to,
\begin{equation}
   \tilde{f}_T = \frac{1}{2\pi}\frac{A_4\sqrt{\tilde{D}_T}(A_1-\tilde{D}_T)}{\tilde{A}_3'\tilde{D}_T+\tilde{A}_2'-A_1\tilde{A}_3'}~,\hspace{1cm}\text{with}\hspace{0.5cm} \tilde{A}'_{\{2,3\}}=\tilde{A}_{\{2,3\}}/\omega\;.
\label{fT_theo}
\end{equation}
The slope of the sigmoid curve at the transition frequency is given by,
\begin{equation}
    \left.\frac{\text{d}\tilde{D}_{sig}}{\text{d}f}\right\rvert_{f=\tilde{f}_T}= R\left[\frac{\tilde D_T}{4 \tilde{f}_T}\text{ln}\left(\frac{\tilde{D}_{high}}{D_{low}}\right)\right]~.
\end{equation}
Due to the complexity of deriving this slope directly from the Guyader model, the parameter $R$ is estimated using a parametric fitting procedure  \cite{arasan_simple_2021-1}. Figure~\ref{Guyader_Arasan} compares on panels~(a) and (b)  the equivalent bending stiffness and the corresponding equivalent wavenumbers respectively predicted by the sigmoid-based model with those obtained from the reference Guyader model. The material and geometric parameters correspond to the three-layer configuration described in Sec.~\ref{Presentation_system}~~(see Tab.~\ref{tab:material_properties}), and identical to those of the sample used for the experimental analysis. For this comparison, perfect interface conditions are assumed ($B=0$ m.Pa$^{-1}$). The results show that the sigmoid-based model accurately reproduces the frequency evolution of both the bending stiffness and the associated equivalent wavenumber predicted by the reference model, with a maximum relative error estimated at $1.2\%$.

\section{EXPERIMENTAL ANALYSIS }
This section presents the methodology used to estimate the equivalent interface parameter under varying excitation amplitudes, based on experimental measurements and the sigmoid-based model. First, the experimental setup employed to excite the structure and record the vibrational response of the three-layer sample with unknown interfacial properties is described. Secondly, the bending contribution is extracted from the measured transverse displacement using the sigmoid model presented in the previous section. Finally, a fitting procedure based oin the sigmoid model is applied to estimate the interface parameter $B$. Its evolution with the excitation amplitude provides a global insight into the variation of the equivalent Young’s modulus, which is then compared with the predictions of the Guyader model using the CFAT method. 

\subsection{Experimental set-up}
\begin{figure*}[ht!]
 \centering
\includegraphics[width=1\linewidth]{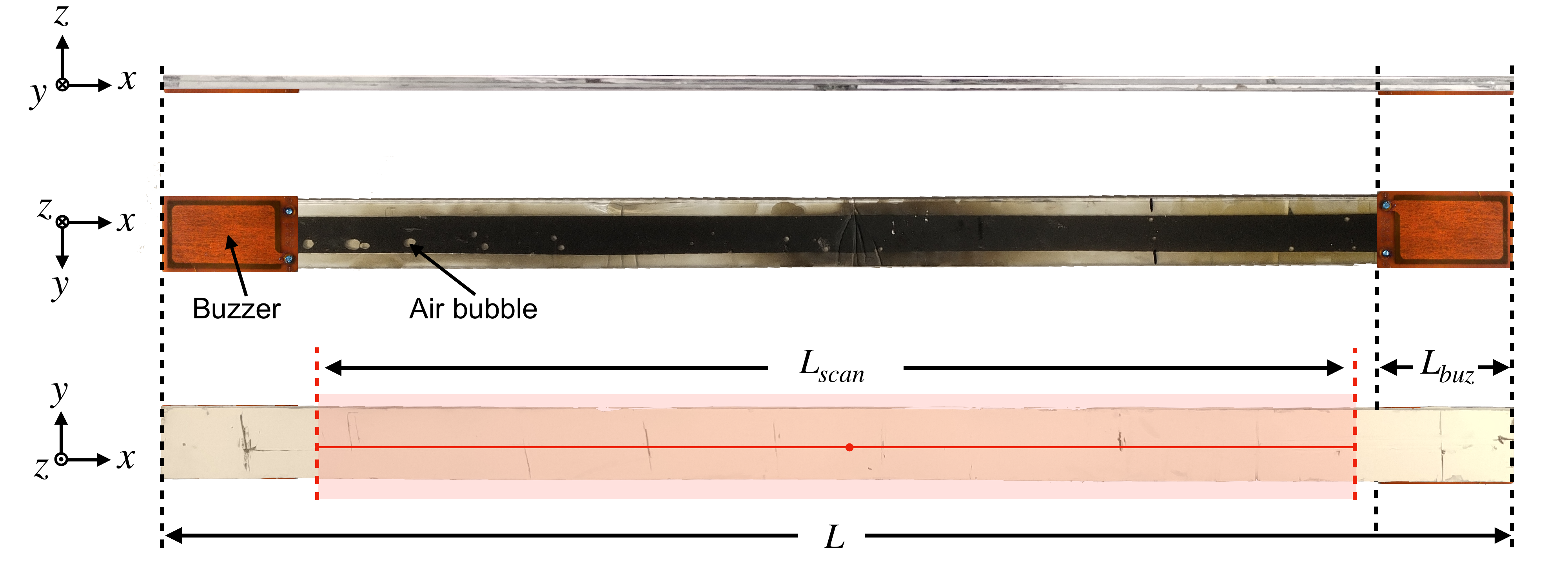}
\vspace{-1cm}
\caption{Lateral, bottom, and top views of the experimental glass/epoxy/glass three-layer beam of length $L$ with imperfect interfaces, on which piezoelectric buzzers of length $L_{buz}$ are bonded. The red line denotes the scanning path of the laser vibrometer used to measure the transverse displacement. To improve measurement quality, the measured surface is covered with a thin white adhesive layer, ensuring an opaque and matte surface.} 
 \label{exp_set_up}
\end{figure*}

The sample beam has a total length of $L=60$ cm and is freely suspended from a supporting frame. Excitation is provided by two piezoelectric buzzers (P-876.A15 SN 63/82), each with a length $L_{buz}=6$ cm, mounted at both ends of the sample see Fig.~\ref{exp_set_up}. Linear frequency chirps are used as input signals, covering a bandwidth of $10$~kHz over the frequency range from $1$~kHz to $50$~kHz, with a resolution of $\text{d}f=6.25$~Hz. The transverse displacement field of the beam,  free of any excitation, is acquired using a scanning laser vibrometer PSV-400 Sc. Head. Measurements are performed along a single line of length $L_{scan}=47.22$~cm, aligned with the longitudinal $x$-axis of the beam and centered at mid-span. The spatial sampling interval is approximately $\text{d}x=2.9$~mm, resulting in $n_x = 163$ measurement points along the structure. At each scan position, the recorded signal is averaged over ten acquisitions in order to improve measurement accuracy.
The amplitude of the excitation signal, denoted $A_0$, is controlled by the Polytech laser vibrometer software and varies from $0$V and $3$V. Seven distinct amplitude levels, listed in Tab.~\ref{tab:A_0_and_A}, are selected within this range to record the different data sets. It should be noted that $A_0$ corresponds to an instrumental input value, influenced by the vibrometer software, an amplifier, and the frequency-dependent response of the piezoelectric buzzers. As such, it does not directly represent the actual vibration level of the beam. To provide a physically meaningful characterization of the structural response, a mechanical amplitude $A$ is introduced by Eq.~(\ref{mechanical_amp}). This quantity is derived from the measured displacement field and provides a global indicator of the vibration level, obtained by averaging over all scan positions and over the frequency range,

\begin{equation}
A = \frac{1}{n_f n_x} \sum_{i=1}^{n_x} \sum_{j=1}^{n_f} \left| u(x_i, f_j) \right|
\label{mechanical_amp}
\end{equation}
where $u(x_i, f_j)$ denotes the measured transverse displacement at position $x_i$ and frequency $f_j$ while $n_x$  ($n_f$) is the number of spatial measurement points (frequency samples). 
Now, each data set is associated with an average mechanical amplitude $A$ (see Tab.~\ref{tab:A_0_and_A}).

\begin{table}[h!]
\centering
\begin{tabular}{|l|ccccccc|}
\hline
Data set & $1$ & $2$ & $3$ & $4$ & $5$ & $6$ & $7$ \\ 
\hline
$A_0$ [V] & $0.1$ & $0.5$ & $1.0$ & $1.5$ & $2.0$ & $2.5$ & $3.0$ \\ 
$A$ [pm] & $22$ & $92$ & $152$ & $198$ & $232$ & $265$ & $290$ \\ 
\hline
\end{tabular}
\vspace{0.3cm}
\caption{Linear chirp excitation levels $A_0$ and their corresponding average mechanical amplitudes $A$ measured on the sample.}
\label{tab:A_0_and_A}
\end{table}
After acquiring the transverse displacement fields of the sample at various excitation levels, the next step is to isolate the contribution of the longitudinal bending mode from the remaining components of the displacement field.
\subsection{Bending wavenumber isolation using the sigmoid-based model}
\label{bending_k_isolation}
\begin{figure}[ht!]
\centering
\includegraphics[width=1\linewidth]{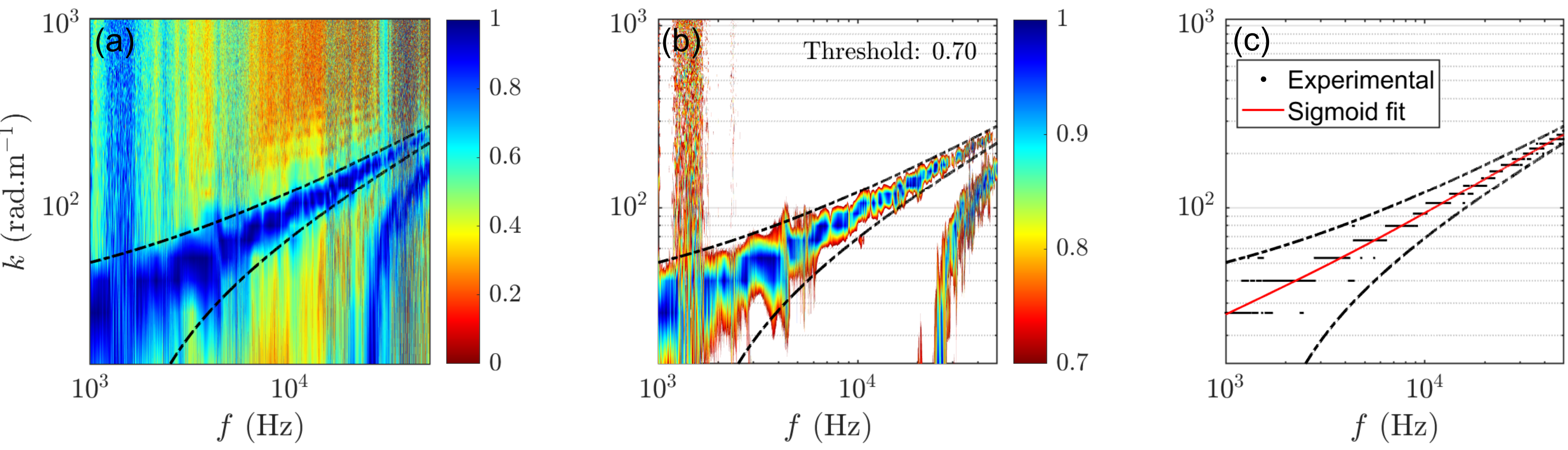}
\vspace{-1cm}
\caption{Procedure to isolate the longitudinal bending wavenumber from the measured displacement field (applied as example to data set 1). Panel~(a) shows the normalized 2DFFT of the displacement fields. In panel~(b), two distinct branches are visible,} and a confidence area is selected to made the parameter fitting on threshold values $\in [0.9;1.0]$ see panel~(c).
 \label{disp_curves}
\end{figure}

To estimate the bending wavenumber $k_b$ at each temporal frequency, the system's dispersion relation is constructed from the recorded displacement field. A spatial FFT is applied to the $n_x$ measurement points, spanning the wavenumber domain $k=\left[\frac{-\pi}{\text{d}x};\frac{\pi}{\text{d}x}\right]$. Considering the scanned length $L_{scan}$, the resulting wavenumber resolution is $\text{d}k = 2\pi/L_{scan} \approx 13.3~\text{rad.m}^{-1}$. Figure~\ref{disp_curves}(a) presents the normalized 2DFFT $(f,k)$ of the measured transverse displacement field for data set 1, revealing two predominant deformation branches. The branch  with the higher wavenumber $k$ corresponds to the longitudinal bending mode propagating along the beam's length ($x$-axis in Fig.~\ref{exp_set_up}), while the second branch is associated with higher order bending mode along the width direction ($y$-axis in Fig.~\ref{exp_set_up})\cite{margerit2018these}. At each frequency, the bending wavenumber $k_{b}$ is extracted from the normalized 2DFFT $(f,k)$ by selecting all components of the wavenumber with amplitudes above a defined threshold ($0.9$ in the present study) within a confidence interval surrounding the branch of interest. This procedure suppresses measurement noise at low frequencies and removes contributions from the secondary mode at higher frequencies, improving the accuracy of the $k_b$ estimation. Finally, a regression is applied to obtain a smooth curve of $k_b$ as a function of frequency. The regression is performed by fitting the cloud of experimental data points in the frequency–wavenumber plane with a curve of the following form,
\begin{equation}
    k_{fit}(f)=\sqrt{2\pi f}D_{sig}(f,\tilde{f}_T,R)^{-1/4}~,
    \label{k_fit}
\end{equation}
       where $\tilde{f}_T$ and $R$ are obtained in the least squares sense and the others parameters are fixed by the analytical model described in Sec.~\ref{sec:sigmoid}~. Figure~\ref{disp_curves} illustrates both the normalized 2DFFT and the method used to isolation of the bending wavenumbers. The cloud of points resulting from 2DFFT maximization at each frequency, as well as the regression curve, are drawn in panel~(c).
       
This procedure is repeated for all seven experimental data sets, and the parameters extracted from the fits using Eq.~(\ref{k_fit}) are listed in Tab.~\ref{tab:ft_and_R}.
\begin{table}[ht!]
\centering
\begin{tabular}{|l|ccccccc|}
\hline
Data set & $1$ & $2$ & $3$ & $4$ & $5$ & $6$ & $7$ \\ 
\hline
$f_T^{exp}$ [kHz] & $19.59$ & $18.90$ & $17.72$ & $17.27$ & $17.33$ & $17.09$ & $16.73$ \\ 
$R$ [-] & $1.10$ & $1.05$ & $1.03$ & $0.99$ & $0.97$ & $0.95$ & $0.95$\\ 
\hline
\end{tabular}
\label{table:ft_R}
\vspace{0.3cm}
\caption{Transition frequency ($f_T$) and slope factor ($R$) obtained by the parametric fitting process.}
\label{tab:ft_and_R}
\end{table}
As the excitation amplitude increases, a general decrease in both $f_T^{exp}$ and $R$ is observed, indicating the onset of nonlinear behavior. The decrease of the slope factor $R$ does not significantly alter the overall shape of the sigmoid curve. The evolution of the transition frequency $\tilde{f}_T$ is more straightforward to interpret, since it can be expressed analytically via the sigmoid model cf. Eq.~(\ref{fT_theo}). This analytical relation allows the equivalent interface parameter $B$ to be estimated for each experimental data set.

\subsection{Interface parameter estimation and nonlinear behavior of the equivalent dynamic Young's modulus}
\begin{figure}[ht!]
 \centering
\includegraphics[width=1\linewidth]{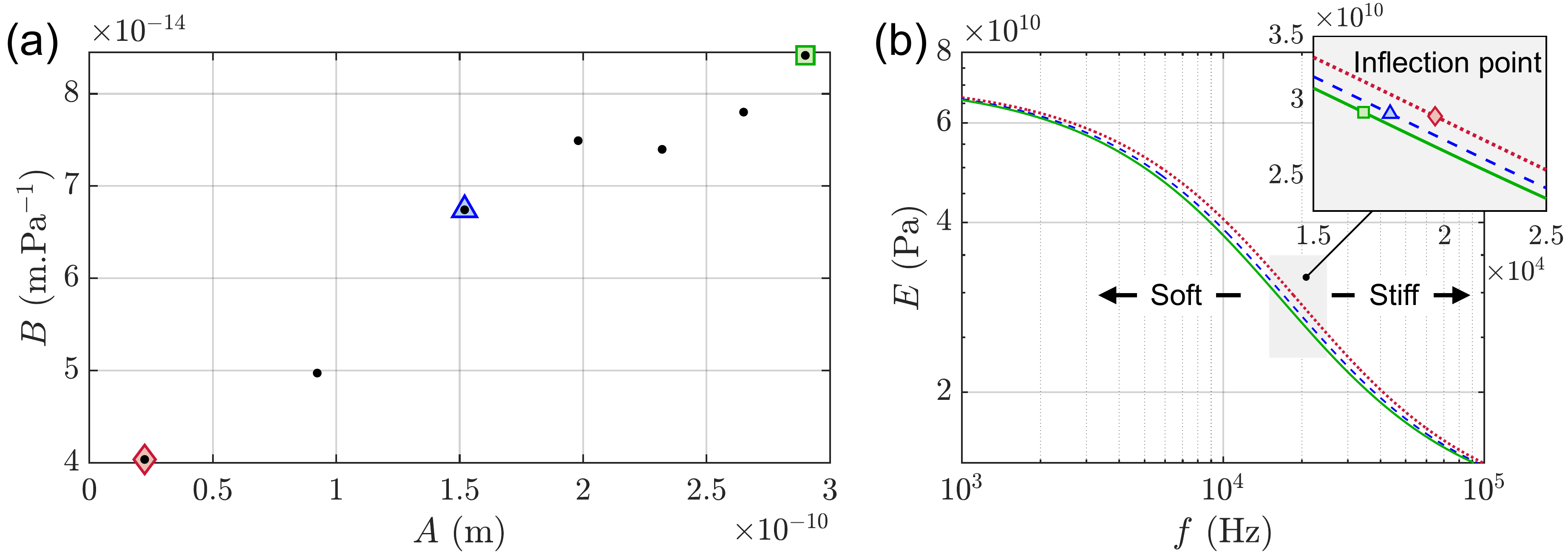}
\vspace{-1cm}
\caption{Influence of the interface parameter $B$ on the dynamic equivalent properties of a three-layer structure from Guyader equivalent plate model. In panel~(a) is displayed the interface parameter as function of the excitation amplitude while in panel~(b) is shown its influence on the analytical prediction. Each symbol \{$\Diamond$, $\triangle$, $\square$\} corresponds to a specific value of $B =\{ 4.04, 6.74, 8.42\}\times$$10^{-14}~($m.Pa$^{-1})$.} 
 \label{B_results}
\end{figure}

The equivalent interface parameter $B$ is determined using a numerical root-finding procedure. Starting from the theoretical expression of the transition frequency $\tilde{f_T}(B)$ write in Eq.~(\ref{fT_theo}), the value of $B$ is obtained by numerically solving the implicit equation $\tilde{f_T}(B)\!=\!f_T^{exp}$, where $f_T^{exp}$ represents the experimentally measured transition frequency. Applying this procedure across all seven data sets enables the investigation of the amplitude dependence of $B$ as reported in Fig.~\ref{B_results}(a).

The results highlight a nonlinear evolution of the interface parameter, with a dominant pattern indicating that $B$ increases with the excitation level. Increasing $B$ directly affects the equivalent dynamic bending stiffness $D(f)$ or the Young’s modulus $E(f)$ of the system which are related by,
\begin{equation}
    D =  \frac{Eh_{eq}^3}{12(1-\nu_{eq}^2)} ~, \hspace{1cm} \text{with}~~h_{eq}=\sum_{n=1}^3 h_n~,~\hspace{1cm}  \text{and}~~\nu_{eq} = \sum_{n=1}^3\frac{\nu_n h_n}{h_{eq}}~.
\end{equation}

As illustrated in Fig.~\ref{B_results}(b), which displays the analytical prediction of the frequency-dependent Young’s modulus obtained from the Guyader model with imperfect interface conditions, increasing the interface parameter induces a downward shift of the transition region between the two asymptotic regimes, indicating a softening of the system at these frequencies. In the low-frequency domain, no differences are expected, in agreement with the prediction given by $D_{low}$. In contrast, in the high-frequency regime, a slight variation of $E$ is theoretically expected, as predicted by Eq.~(\ref{D_high}). The variation of $B$ between the lowest and highest excitation amplitudes spans a factor of two, leading to significant changes in $E$ over a broad frequency range covering nearly two decades. The frequency shift at the transition frequency is estimated to be $\Delta \tilde{f_T} \approx 3~\text{kHz}$, as reported in Table~\ref{table:ft_R}.
\newline

To allow for comparison between the model (with the estimated $B$) and the experimental measurements, the CFAT methodology (Corrected Force Analysis Technique) is used to evaluate the equivalent dynamic bending stiffness of the structure \cite{ege_assessment_2018,madinier_spatial_2025} and to analyze the evolution of the frequency shift associated with its sigmoidal behavior. This method consists in using the measured transverse displacement fields on a given mesh grid and injecting it into the local equation of motion discretized by finite difference schemes (in our case a flexural beam) \cite{leclere_vibration_2012,leclere_practical_2015}. Using a least squares approach, the estimated structural parameter over multiple points across the sample can be used to provide a frequency-dependent equivalent value.  
\begin{figure}[ht!]
 \centering
\includegraphics[width=1\linewidth]{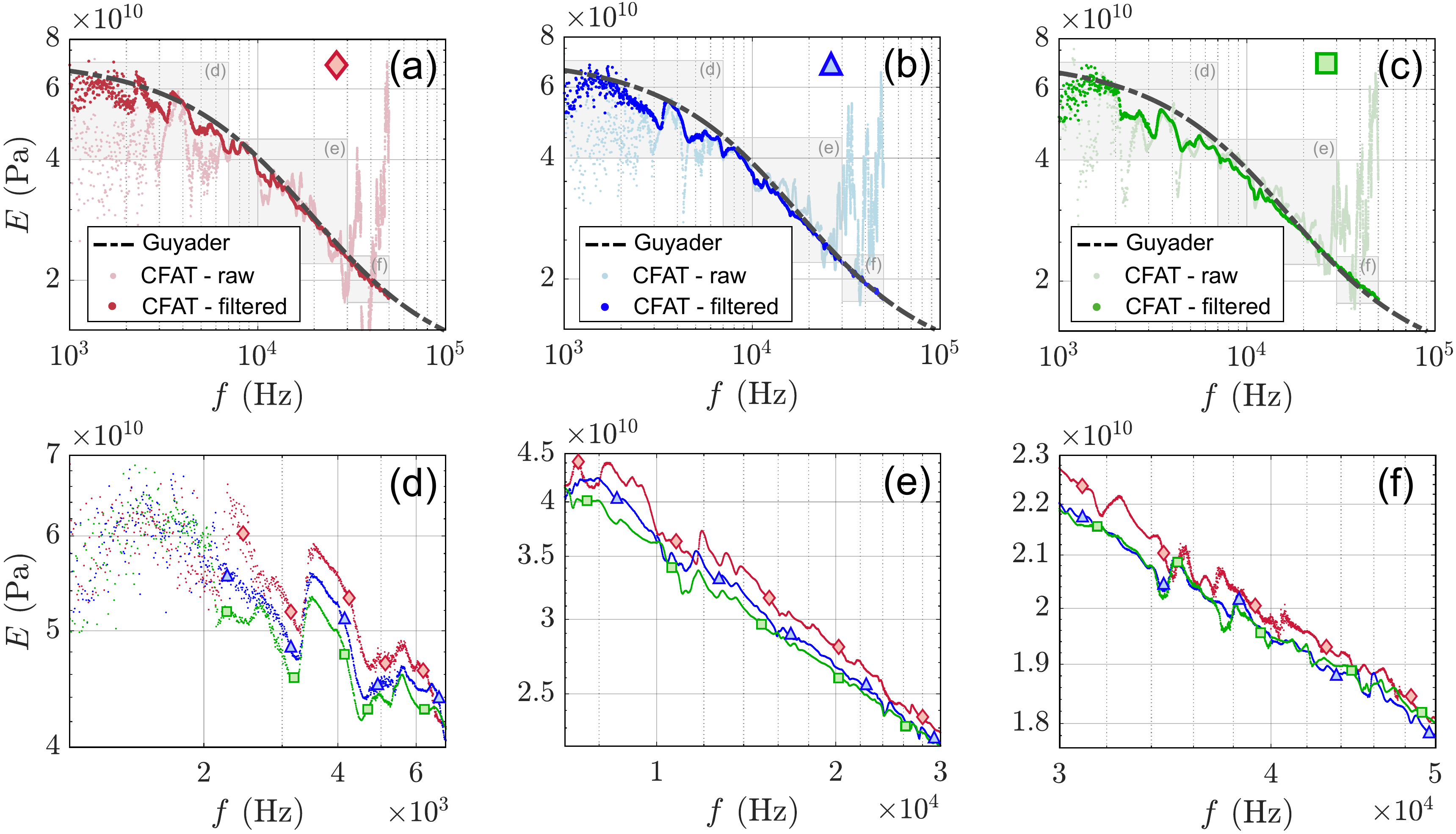}
\vspace{-1cm}
\caption{Panels~(a, b, c) compare the dynamically equivalent Young’s modulus obtained from experimental data using the CFAT method (with and without the filtering procedure), to the estimated analytical predictions (Guyader model) for $B =\{ 4.04, 6.74, 8.42\}\times$$10^{-14}~($m.Pa$^{-1})$. Panels~(d,~e,~f) show the frequency shift of the equivalent dynamic Young’s modulus obtain (CFAT - filtered) for different frequency bandwidths.}
 \label{Freq_shift}
\end{figure}

Fig.~\ref{Freq_shift}(a, b, c) shows the equivalent Young’s modulus obtained using the CFAT method (CFAT~-~raw), directly compared to the predictions presented in Fig.~\ref{B_results}(b). For clarity, only three values are shown, corresponding to the data sets 1 ($\Diamond$), 3 ($\triangle$) and, 7 ($\square$). The CFAT results exhibit good agreement with the predictions in the frequency range around $7$ kHz to $30$ kHz. Below this frequency band, various factors, including measurement noise, the frequency response of the buzzers, and the finite dimension of the sample, may contribute to an inaccurate estimation of the wavenumber and therefore of the structure parameter. As frequency increases, the sample’s response tends to be smoother and the number of spatial periods increases, leading to a better estimation of the bending wavenumber. In contrast, at higher frequencies, a significant increase of $E$ is observed and can be attributed to the onset of an additional deformation mode, as seen by the second branch in Fig.~\ref{disp_curves}(a, b). To limit this influence, the combination of filtering and border-padding approach has been developed \cite{auquier_imperfect_2024}, leading to a significant improvement in high-frequency estimation, as can be seen through the "CFAT~-~filtered" curves. Using the CFAT–filtered curves, the nonlinear behavior of the material properties becomes more easily observable across the lower, middle, and higher frequency bands see Fig.~\ref{Freq_shift}(d, e, f). As expected, the frequency shift is more pronounced near the transition frequency due to the sigmoid evolution of $E(f)$.

\section{CONCLUSION}
In this work we have show that, mechanical multilayer structure assembled by gluing layers with manufacturing-induced defects, exhibits imperfect interfaces between the core and the skins with nonlinear behavior. These nonlinear interface properties lead to changes in the structure’s equivalent material properties as a function of the excitation level. In particular, a global shift of the equivalent bending stiffness toward lower frequencies is observed as the excitation level increases, indicating a softening of the system.

 The main contribution of this work is the experimental identification of an equivalent nonlinear interface parameter $B$, identified from vibrational response measurements on a glass/epoxy/glass three-layer beam at different excitation amplitudes. This parameter was determined by fitting the experimental data with a sigmoid function and by tracking the frequency shift of the transition frequency, which serves as an indicator of the interface nonlinearity behavior. Over the investigated range of excitation levels, the identified interface parameter was found to vary by approximately a factor of two, leading to a significant modification of the equivalent dynamic bending stiffness across a wide frequency range. This result is particularly noteworthy and promising, as such amplitude-dependent variations in equivalent bending properties linked to imperfect interfaces have, to our knowledge, not yet been documented.
 
To extend the present findings, future work will focus on increasing the number of tested samples with varying bonding qualities, in order to investigate the evolution and magnitude of the interface parameter as a function of bonding quality. In addition, the development of multilayer samples produced using additive manufacturing techniques such as 3D printing~\cite{ngo_additive_2018} would enable a much finer control over the nature and distribution of induced heterogeneities. Such an approach would also make it possible to design interface geometries that are either passively or actively controlled through the integration of transducers between the layers, thereby enabling tunable equivalent mechanical properties of the multilayer structure.

\section*{ACKNOWLEDGEMENTS}
This work has been done during the postdoctoral position of the first author at the LVA (INSA Lyon) and LTDS (ENTPE) financed by LabEx CeLyA (ANR-10-LABX-0060) of Université de Lyon, within the program “Investissements d’Avenir” operated by the French National Research Agency (ANR). The authors acknowledge Fabien Marchetti, Fabien Chevillotte and François-Xavier Bécot (research engineers at Matelys - Research Lab) for fruitful discussions and valuable scientific advice. They also thank Julien Chatard (technician at LVA) and Théodore Braule (PhD student at LVA, LaMCoS and Saint-Gobain Recherche Paris) for their valuable assistance with the setup and execution of the experimental measurements. 

\appendix
\newpage
\section{Transfer matrix details}
\label{Appendix_A}
The purpose of this appendix is to detail the derivation steps leading to the Zig–Zag model employed in this work. In particular, it aims express the displacement field of an arbitrary layer $n$ of the structure in terms of that of the first layer, which is chosen as the reference layer. The first step of the derivation consists in relating the stress field to the displacement field through the strain field. As mentioned in Sec.~\ref{Presentation_system}, each layer $n=\{1,2,3\}$ is assumed to be isotropic. Accordingly, the constitutive law relating stress and strain fields is defined by the following stiffness tensor,

\begin{equation}
\begin{array}{c@{\qquad}l}
\begin{pmatrix}
    \sigma^n_{xx} \\ 
    \sigma^n_{yy} \\ 
    \sigma^n_{yz} \\
    \sigma^n_{xz} \\
    \sigma^n_{xy}
\end{pmatrix}
=
\begin{pmatrix}
    Q^n_{11} & Q^n_{12} & 0 & 0 & 0 \\
    Q^n_{12} & Q^n_{22} & 0 & 0 & 0 \\
    0 & 0 & Q^n_{44} & 0 & 0 \\
    0 & 0 & 0 & Q^n_{55} & 0 \\
    0 & 0 & 0 & 0 & Q^n_{66}
\end{pmatrix}
\begin{pmatrix}
    \epsilon^n_{xx} \\ 
    \epsilon^n_{yy} \\ 
    2\epsilon^n_{yz} \\
    2\epsilon^n_{xz} \\
    2\epsilon^n_{xy}
\end{pmatrix}, \hspace{0.5cm}\text{with}
&
\begin{alignedat}{2}
&Q^n_{11} = Q^n_{22} = \frac{E_n}{1-\nu_n^2}~,\\
&Q^n_{12} = \frac{\nu_n E_n}{1-\nu_n^2}~,\\
&Q^n_{44} = Q^n_{55} = Q^n_{66}= \frac{E_n}{2(1+\nu_n)}~.
\end{alignedat}
\end{array}
\end{equation}

Using the multilayer displacement field in Eq.~(\ref{Champ_des_deplacements}), the strain field of the $n$th layer in the global reference coordinate system is expressed as,

\begin{equation}
  \begin{pmatrix}
      \epsilon^n_{xx} \\
      \epsilon^n_{yy} \\
      2\epsilon^n_{xy} \\
      2\epsilon^n_{xz} \\
      2\epsilon^n_{yz} \\
  \end{pmatrix}=
    \begin{pmatrix}
    \frac{\partial}{\partial x} & 0 & 0 \\
    0 & \frac{\partial}{\partial y} & 0 \\
    \frac{\partial}{\partial y} & \frac{\partial}{\partial x} & 0 \\
    \frac{\partial}{\partial z} & 0 & \frac{\partial}{\partial x} \\
    0 & \frac{\partial}{\partial z} & \frac{\partial}{\partial y} \\
  \end{pmatrix} \begin{pmatrix}
      u^n_x \\
      u^n_y \\
      u^n_z \\
  \end{pmatrix}=\begin{pmatrix}
      \frac{\partial u^n_x }{\partial x} \\
      \frac{\partial u^n_y }{\partial y}
      \\ 
      \frac{\partial u^n_x }{\partial y}+\frac{\partial u^n_y }{\partial x}\\
      \frac{\partial u^n_x }{\partial z}+\frac{\partial u^n_z }{\partial x} \\
      \frac{\partial u^n_y }{\partial z}+\frac{\partial u^n_z }{\partial y}
  \end{pmatrix}~.
\end{equation}
The component $\epsilon^n_{zz}$ vanishes, since the assumption of constant transverse displacement implies no deformation across the thickness. Using the displacement field definition provided in Eq.~(\ref{Champ_des_deplacements}), the governing equations (Eq.~(\ref{displacements_conditions})) can be expressed as follows,

\begin{equation}
 \left \{
    \begin{split}
        \psi^n_x+\frac{h_n}{2}\left[\frac{\partial W}{\partial x}+\phi^n_x\right]=\psi^{n-1}_x-\frac{h_{n-1}}{2}\left[\frac{\partial W}{\partial x}+\phi^{n-1}_x\right] +\hat{u}^{n-1}_x \\
        \psi^n_y+\frac{h_n}{2}\left[\frac{\partial W}{\partial y}+\phi^n_y\right]=\psi^{n-1}_y-\frac{h_{n-1}}{2}\left[\frac{\partial W}{\partial y}+\phi^{n-1}_y\right]+\hat{u}^{n-1}_y \\
    \end{split}
    \right.~,
\label{Continuite_Champ_des_deplacements}
\end{equation}
while the continuity conditions for shear stresses Eq.~(\ref{shear_stresses_conditions}) are written as,
\begin{equation}
 \left \{
    \begin{split}
        Q^n_{55}2\epsilon^{n}_{xz}  &= Q^{n-1}_{55}2\epsilon^{n-1}_{xz}\,  \\
        Q^n_{44}2\epsilon^{n}_{yz} &= Q^{n-1}_{44}2\epsilon^{n-1}_{yz}\, 
    \end{split}
    \right., \hspace{1cm}\text{with}\,,\hspace{0.5cm}
  \begin{pmatrix}
      2\epsilon^n_{xz} \\
      2\epsilon^n_{yz} \\
  \end{pmatrix}=
\begin{pmatrix}
      \frac{\partial u^n_x }{\partial z}+\frac{\partial u^n_z }{\partial x} \\
      \frac{\partial u^n_y }{\partial z}+\frac{\partial u^n_z }{\partial y}
  \end{pmatrix}=
  \begin{pmatrix}
      -\cancel{\frac{\partial W}{\partial x}}-\phi^n_x+\cancel{\frac{\partial W}{\partial x}}\\
       -\cancel{\frac{\partial W}{\partial y}}-\phi^n_y+\cancel{\frac{\partial W}{\partial y}}\\
  \end{pmatrix}\,,
\end{equation}
and finally,

\begin{equation}
 \left \{
    \begin{split}
        Q^n_{55}\phi^n_x  &= Q^{n-1}_{55}\phi^{n-1}_x \\
        Q^n_{44}\phi^{n}_y &= Q^{n-1}_{44}\phi^{n-1}_y
    \end{split}
    \right. \; .
\label{Continuite_contraintes}
\end{equation}
Since each condition in Eqs.~(\ref{Continuite_Champ_des_deplacements}–\ref{Continuite_contraintes}) is now expressed in terms of the kinematic variables of the associated layer, they can be collected and written in the following matrix form,
\begin{equation}
    \begin{pmatrix}
    1 & 0 & 0 \\
    0 & Q_{55}^n& 0 \\
    h_n/2 & h_n/2  & 1 \\
  \end{pmatrix}\begin{pmatrix}
        \frac{\partial W}{\partial x} \\
        \phi^n_x\\
        \psi^n_x\\\end{pmatrix} =  \begin{pmatrix}
    1 & 0 & 0 \\
    0 & Q_{55}^{n-1}& 0 \\
    -h_{n-1}/2 & -h_{n-1}/2-B Q^{n-1}_{55}  & 1 \\
  \end{pmatrix}\begin{pmatrix}
        \frac{\partial W}{\partial x} \\
        \phi^{n-1}_x\\
        \psi^{n-1}_x\\\end{pmatrix}-\begin{pmatrix}
        0 \\
       0 \\
        B_0\\\end{pmatrix}\;.
\end{equation}
leading to,
\begin{equation}
\underbrace{
\begin{pmatrix}
      \frac{\partial W}{\partial x} \\
      \phi^n_x \\
      \psi^n_x \\
      \frac{\partial W}{\partial y} \\
      \phi^n_y \\
      \psi^n_y
\end{pmatrix}
}_{\{L^n\}}
=
\underbrace{
\begin{pmatrix}
    1 & 0 & 0 & 0 & 0 & 0  \\
    0 & \mathcal{A}^n_{x} & 0 & 0 & 0 & 0  \\
    \mathcal{B}^n & \mathcal{C}^n_{x} & 1 & 0 & 0  & 0  \\
    0 & 0 & 0 & 1 & 0 & 0  \\
    0 & 0 & 0 & 0 & \mathcal{A}^n_{y}  & 0   \\
    0 & 0 & 0 & \mathcal{B}^n & \mathcal{C}^n_{y} & 1
\end{pmatrix}
}_{[T^n]}
\underbrace{
\begin{pmatrix}
      \frac{\partial W}{\partial x} \\
      \phi^{n-1}_x \\
      \psi^{n-1}_x \\
      \frac{\partial W}{\partial y} \\
      \phi^{n-1}_y \\
      \psi^{n-1}_y
\end{pmatrix}
}_{\{L^{n-1}\}}-\underbrace{
\begin{pmatrix}
      0\\
      0 \\
      B_0 \\
      0 \\
      0 \\
      B_0
\end{pmatrix}
}_{\{B_0\}} \;,
\label{Matrix_T_appendix}
\end{equation}
where the coefficients are defined as,
\begin{equation}
    \mathcal{A}^n_x = \frac{Q_{55}^{n-1}}{Q_{55}^{n}}\;,\hspace{1cm} \mathcal{B}^n = - \frac{h_n+h_{n-1}}{2}\;, \hspace{1cm} \mathcal{C}^n_x =-\frac{h_n}{2}\mathcal{A}^n_x-\frac{h_{n-1}}{2}-BQ_{55}^{n-1}\;.
\end{equation}

By recursively applying the relationship written in Eq.~(\ref{Matrix_T_appendix})
 from the outermost layer $N$ back to the first one, 
 \begin{equation}
     \{L^n\}=[T^n][T^{n-1}][T^{n-2}]\cdots[T^2]\{L^{1}\}-(n-1)\{B_0\}\;, \hspace{1cm}\text{with}\,,\hspace{0.5cm} n=\{2,\ldots,N\}\;,
 \end{equation}
 the displacement of each layer can be expressed as functions of the kinematic variables of the first layer, using the following matrix,

\begin{equation}
[\mathcal{T}^n]=\begin{pmatrix}
    1 & 0 & 0 & 0 & 0 & 0  \\
    0 & \alpha^n_{x} & 0 & 0 & 0 & 0  \\
     \beta^n & \gamma^n_{x} & 1 & 0 & 0  & 0  \\
    0 & 0 & 0 & 1 & 0 & 0  \\
    0 & 0 & 0 & 0 & \alpha^n_{y}  & 0   \\
    0 & 0 & 0 & \beta^n & \gamma^n_{y} & 1   \\
\end{pmatrix}
\hspace{1cm}\text{leading to}\;, \{L^n\}=[\mathcal{T}^n]\{L^{1}\}-(n-1)\{B_0\}\;.
\label{LnL1}
\end{equation}
The coefficients $\alpha, \beta, \gamma$ depend on the geometric and mechanical properties of each layer as follows,
\begin{subequations}
\begin{align}
\beta^n 
&= \sum_{k=1}^n \mathcal{B}^k 
= \sum_{k=1}^n -\frac{h_{k-1}+h_k}{2} 
= R_1 - R_n \;, 
\label{betan} \\
\alpha_{x}^n 
&= \prod_{i=1}^{n} \mathcal{A}_{x}^i 
= \frac{C_{55}^1}{C_{55}^n} \;, \\
\alpha_{y}^n 
&= \prod_{i=1}^{n} \mathcal{A}_{y}^i 
= \frac{C_{44}^1}{C_{44}^n} \;, \\
\gamma_{x}^n 
&= \sum_{k=2}^n \left(\prod_{i=1}^{k-1} \mathcal{A}_{x}^i\right)\mathcal{C}^k_{x} = -\frac{h_1}{2} - (n-1) Q_{55}^1 B - \sum_{k=2}^n \frac{Q_{55}^1}{Q_{55}^k} \frac{h_k}{2} \;, \\
\gamma_{y}^n 
&= \sum_{k=2}^n \left(\prod_{i=1}^{k-1} \mathcal{A}_{y}^i\right)\mathcal{C}^k_{y} = -\frac{h_1}{2} - (n-1) Q_{44}^1 B - \sum_{k=2}^n \frac{Q_{44}^1}{Q_{44}^k} \frac{h_k}{2} \;.
\label{gammany}
\end{align}
\end{subequations}

The interface parameter $B$ affects only the coefficients $\gamma^n_{\{x,y\}}$. When $B=0$, i.e. in the case of perfect interfaces, the present formulation reduces to the expressions obtained by Guyader and Caciollati \cite{Guyader2007Viscoelastic}. Finally, the kinematic field $\{L^n\}$ can be expressed as a function of $\{L^1\}$ using Eq.~(\ref{LnL1}) leading to the following kinematic field,

 \begin{equation}
      \left \{
      \begin{split}
      u^n_x& = \psi^1_x(x,y,t)+F_\omega \frac{\partial W(x,y,t)}{\partial x}+F^n_{x}\phi^1_x(x,y,t)-(n-1)B_0\\ 
      u^n_y& =\psi^1_y(x,y,t)+F_\omega \frac{\partial W(x,y,t)}{\partial y}+F^n_{y}\phi^1_y(x,y,t)-(n-1)B_0\\ 
      u^n_z& =W(x,y,t) \\ 
            \end{split}
      \right.\; ,
      \label{champ_depl_n_fonction_1_annexe}
  \end{equation}
with, 
\begin{equation}
    F^n_{\{x,y\}}=\alpha^n_{\{x,y\}}(R_n-z)+\gamma^n_{\{x,y\}}~, \hspace{1cm} \hspace{1cm} \text{and}\,, \hspace{1cm} F_\omega = R_1-z~.
    \label{coeff_F_annexe}
\end{equation}

\newpage
\bibliography{References}
\bibliographystyle{unsrt}
\end{document}